\documentclass[aps,prd,amsmath,amssymb,twocolumn,preprintnumbers,nofootinbib]{revtex4-1}
\pdfoutput=1
\usepackage[utf8]{inputenc} 
\usepackage{graphicx}
 \usepackage{multirow}
\graphicspath{{./figures/}}
\usepackage{url}
\usepackage[bookmarks, pagebackref=false]{hyperref}
\usepackage[usenames,dvipsnames]{xcolor}
\definecolor{orange}{cmyk}{0,0.5,1,0}
\definecolor{rossoCP3}{cmyk}{0,.88,.77,.40}
\definecolor{graa}{rgb}{0.8,0.8,0.8}
\definecolor{blaa}{rgb}{0.2,0.2,0.6}
		\hypersetup{
			colorlinks, 
			bookmarksopen, 
			bookmarksnumbered,
			citecolor=blaa, 		
			linkcolor=rossoCP3,	
			urlcolor=rossoCP3,			
			}
\usepackage{amsthm}
\usepackage{bm}
\usepackage{bbm}
\usepackage{pxfonts}

\usepackage{amsmath,amssymb,amsfonts}
\usepackage{color}
\usepackage{float}
\usepackage{hyperref}
\usepackage[Symbolsmallscale]{upgreek}
\usepackage{amsmath}
\usepackage{amsfonts}
\usepackage{amssymb,dsfont}
\usepackage{graphicx}
\usepackage{amssymb}
\usepackage[vcentermath]{youngtab}
\usepackage[all]{xy}
\usepackage{pstricks}
\usepackage{dsfont}%
\usepackage{placeins}
\usepackage{xspace}
\usepackage{cancel} 

\usepackage{slashed}
\usepackage[caption=false, labelformat=simple, listofformat=subsimple, labelfont=default, margin=5pt, justification=raggedright]{subfig}

	\makeatletter
		\renewcommand{\p@subfigure}{}
	\makeatother

\usepackage{natbib}

\usepackage{feynmf}

\newcommand{\beq}{\begin{eqnarray}}
\newcommand{\eeq}{\end{eqnarray}}

\newcommand{\bmp}{\noindent\begin{minipage}{16cm}}
\newcommand{\emp}{\end{minipage}\vskip 7mm} 

\def\lsim{\mathrel{\rlap{\lower4pt\hbox{\hskip1pt$\sim$}}
    \raise1pt\hbox{$<$}}}                
\def\gsim{\mathrel{\rlap{\lower4pt\hbox{\hskip1pt$\sim$}}
    \raise1pt\hbox{$>$}}}                

\newcommand{\brakets}[1]{\left\langle #1 \right\rangle}
\newcommand{\dd}{\mathop{}\!\mathrm{d}}

\renewcommand{\div}{\mathop{}\!\mathrm{div}}

\DeclareFontFamily{U}{mathx}{\hyphenchar\font45}
\DeclareFontShape{U}{mathx}{m}{n}{<-> mathx10}{}
\DeclareSymbolFont{mathx}{U}{mathx}{m}{n}
\DeclareMathAccent{\widebar}{0}{mathx}{"73}

\begin{document}


\title{\texorpdfstring{\Large\color{rossoCP3}  $a$-theorem at  large $N_f$}{}}
\author{Oleg {\sc Antipin}$^{\color{rossoCP3}{\clubsuit}}$}
 \author{Nicola Andrea {\sc Dondi}$^{\color{rossoCP3}{\diamondsuit}}$}
 \author{Francesco {\sc Sannino}$^{\color{rossoCP3}{\diamondsuit}}$}
 \author{Anders Eller {\sc Thomsen}$^{\color{rossoCP3}{\diamondsuit}}$}
\affiliation{\mbox{ $^{\color{rossoCP3}{\clubsuit}}$ Rudjer Boskovic Institute, Division of Theoretical Physics, Bijeni\v cka 54, HR-10000 Zagreb, Croatia}\\\mbox{{ $^{\color{rossoCP3}{\diamondsuit}}$\color{rossoCP3} {CP}$^{ \bf 3}${-Origins}} \&  {\color{rossoCP3}\rm{Danish IAS}},  University of Southern Denmark, Campusvej 55, Odense M - DK-5230, Denmark}
}

\begin{abstract} 
We determine the Jack and Osborn $a$-function and  related metric for gauge-fermion theories to leading order in the large number of fermions   and to all orders in the gauge coupling, demonstrating that the strong $a$-theorem is violated for the minimal choice of the $ a $-function. \\
[.3cm]
{\footnotesize  \it Preprint: CP$^3$-Origins-2018-028 DNRF90
}
\end{abstract}
\maketitle

Quantum Field Theory (QFT) is the language chosen by nature to describe its fundamental  laws with the renormalization group (RG) flow connecting physics at different energy scales. Remarkably this flow is thought to be irreversible as encompassed by Cardy's proposal of the 4-dimensional $a$ theorem~\cite{Cardy:1988cwa}, originally inspired by the 2-dimensional proof of the Zamolodchikov $c$-theorem \cite{Zamolodchikov:1986gt}. The main idea is that one can, in principle, define a monotonically decreasing function from the ultraviolet (UV) to the infrared (IR) along the flow.  

Our goal is to push forward the state-of-the-art by computing the $a$-function to all orders in the couplings, for nonsupersymmetric field theories, exploiting the large number of flavors limit~\cite{Holdom:2010qs}.

In order to define the relevant quantities we start by considering a generic theory defined via the bare Lagrangian $\mathcal{L}_0(g_{0}^{i}, \phi_{0})$ for a set of fundamental fields denoted by $\phi_{0}$. We restrict ourselves to the case in which the couplings, $g^i_0$, are associated to marginal operators.
The theory is extended to curved background, changing the metric $\eta_{\mu\nu} \rightarrow \gamma_{\mu\nu}(x)$ so that it is classically invariant under $ \text{diff} \, \times\, \text{Weyl} $, 
and to spacetime-dependent couplings $g_0^i \rightarrow g_{0}^{i}(x)$. 

Within the extended theory the (renormalized) couplings act as sources for associated composite operators, $ \mathcal{O}_i = \tfrac{\delta S}{\delta g^{i}} $. If we appropriately renormalize the vacuum energy functional, $W = W[\gamma_{\mu\nu}, g^i ]$, such that it is finite, we can extract renormalized composite operator correlators through functional derivatives of $W$. To achieve this, usual renormalization is insufficient: the  spacetime-dependence of the couplings induces new divergences that have to be canceled by new counter terms (CTs) proportional to coupling derivatives. These of course vanish in the limit of constant sources.

We restrict ourselves to only listing the field-independent CTs. The new Lagrangian in $d=4-\epsilon$ reads~\cite{Jack:1990eb} 
	\begin{equation} \label{eq:counter_terms}
	\widetilde{\mathcal{L}}_0(g_{0}^{i}, \phi_{0}) = \mathcal{L}_0(g_{0}^{i}, \phi_{0}) + \mu^{-\epsilon} \lambda(g) \cdot \mathcal{R}\ .
	\end{equation}
Of all 4-dimensional CTs built from metric and couplings that can appear in $\lambda \cdot \mathcal{R} $ we are mainly interested in the following ones:
	\begin{equation} 
	\begin{split}
	\lambda \cdot \mathcal{R}  \supset &   \lambda_a E_4 
	+ \tfrac{1}{2} \mathcal{G}_{ij}  \partial_\mu g^{i} \partial_\nu g^{j} G^{\mu\nu} + \tfrac{1}{2} \mathcal{A}_{ij} \, \nabla^{2} g^{i} \nabla^2 g^{j} \\
	&\; + \tfrac{1}{2} \mathcal{B}_{ijk} \, \partial_\mu g^{i} \partial^\mu g^{j} \nabla^2 g^{k}  \ , 
	\label{eq:RCT}
	\end{split}
	\end{equation}
where $E_4$ is the Euler density and $ G_{\mu\nu} $ is the Einstein tensor. In the MS scheme, the CT coefficients only contain poles in $ \epsilon $; there are no finite parts. These terms are needed to renormalize specific contact divergences in composite operator correlators, so they can be written directly in terms of the renormalized couplings, $ g^{i} $. 

In this framework, RG transformations are deeply interconnected with Weyl rescalings:
\begin{equation}
\gamma_{\mu\nu} \rightarrow e^{-2\sigma} \gamma_{\mu\nu}\ ,\quad \phi \rightarrow e^{\sigma \Delta_{\phi}} \phi\ .
\end{equation} 
We know that for a general QFT in curved space the Weyl symmetry, when present at the classical level, is anomalous. We take $\{ g^i \}$ to couple to the set of all marginal operators defined at $ g^i =0 $, so there exists a source transformation law such that the field-dependent part of the action is invariant under Weyl transformation~\cite{Osborn:1991gm,Baume:2014rla}. The associated  operator  $\Delta_{\sigma}\equiv \int_{} \dd^d x\,  
\sigma(x)
		[ 2 \gamma^{\mu\nu} \tfrac{\delta}{\delta \gamma^{\mu\nu}}
		- \hat{\beta}^i \tfrac{\delta}{\delta g^i} ]
$ acting on $W$ transforms the field-independent CTs  
	\begin{equation} \label{WeylVariation}
	\begin{split}
	\Delta_{\sigma} W & = \Delta_{\sigma} \int \dd^dx\sqrt{\gamma} \mu^{-\epsilon} \lambda \cdot \mathcal{R} \\
	& =	\int \dd^d x\sqrt{\gamma} \mu^{-\epsilon} (\sigma \beta_{\lambda} \cdot \mathcal{R}+\partial_{\mu} \sigma \mathcal{L}^\mu) \ . 
	\end{split}
	\end{equation}
Here $\mu \tfrac{\dd }{\dd \mu} g^{i} = \hat{\beta}^{i} = - \rho^{i} g^{i} \epsilon + \beta^{i}(g)$~\footnote{$ \rho^{i} $ is defined such that $ g^{i}_0 \mu^{-\rho^{i} \epsilon} $ is dimensionless, and its index does not count in the summation convention.}. The $\sigma$-dependent part of eq. \eqref{WeylVariation} has a tensor expansion analogous to \eqref{eq:RCT}:
	\begin{equation} 
	\begin{split}
	\beta_\lambda \cdot \mathcal{R}  \supset & \, a E_4 
	+ \tfrac{1}{2} \mathcal{\chi}_{ij}^g  \partial_\mu g^{i} \partial_\nu g^{j} G^{\mu\nu} + \tfrac{1}{2} \mathcal{\chi}_{ij}^a \, \nabla^{2} g^{i} \nabla^2 g^{j}  \\
	&\; + \tfrac{1}{2} \mathcal{\chi}_{ijk}^b \, \partial_\mu g^{i} \partial^\mu g^{j} \nabla^2 g^{k}  \ .
	\end{split}
	\end{equation}
Since $\left[\mu \frac{\partial}{\partial \mu} + \int \dd^d x \, 2 \gamma^{\mu\nu} \frac{\delta}{\delta \gamma^{\mu\nu}} \right]W = 0$ by dimensional analysis,
	for a global ($\sigma=$ const) transformation,  the integrand of \eqref{WeylVariation} becomes
	\begin{equation} \label{eq:counter_term_betas}
	\left[\epsilon - \int \dd^d y \, \hat{\beta}^{i} (y) \dfrac{\delta}{\delta g^{i}(y)} \right]\lambda \cdot \mathcal{R} = \beta_\lambda \cdot \mathcal{R} \ .
	\end{equation}
From this we  derive, eg. 
	\begin{align}
	\chi^{g}_{ij} &= \left(\epsilon - \hat{\beta}^{\ell} \partial_\ell \right) \mathcal{G}_{ij} - \mathcal{G}_{\ell j} \partial_i \hat{\beta}^{\ell} -  \mathcal{G}_{i\ell} \partial_j \hat{\beta}^{\ell}\ , \label{eq:chi_g}\\
	\chi^{a}_{ij} &= \left(\epsilon - \hat{\beta}^{\ell} \partial_\ell \right) \mathcal{A}_{ij} - \mathcal{A}_{\ell j} \partial_i \hat{\beta}^{\ell} -  \mathcal{A}_{i\ell} \partial_j \hat{\beta}^{\ell}\ , \label{eq:chi_a}
	\end{align} 
which we will need later. 

Since $\Delta_{\sigma}W$ has to be finite by construction, $\chi^g$ and $\chi^a$ must be finite too. This implies that \eqref{eq:chi_g} and \eqref{eq:chi_a} can be interpreted as RG equations for the CTs $\mathcal{G}_{ij} $ and $ \mathcal{A}_{ij} $.
The same reasoning is independently valid for the term proportional to $\partial_{\mu} \sigma$ in \eqref{WeylVariation}. The constraints originating from the requirement of finiteness\footnote{These relations can be also recovered by imposing the Weyl variation to be Abelian: $[ \Delta_{\sigma_1}, \Delta_{\sigma_2} ] W = 0$. } were found in \cite{Jack:1990eb}, and lead to non-trivial relations between CTs. First off 
\begin{equation} 
	8\partial_i \tilde{a} = \left( \chi_{ij}^g + (\partial_i w_j - \partial_j w_i) \right) \beta^j\ , \qquad 8\tilde{a} \equiv 8a + w_i \beta^i \ ,\label{eq:consistencycondition}
\end{equation}
where $w_i$ is a 1-form parametrizing a renormalization scheme redundancy, and $ \tilde{a} $ coincides with $ a $ at fixed points (FPs). It follows that $\tilde{a}$ satisfies the gradient flow equation
\begin{equation}
	8\mu \frac{d\tilde{a}}{d\mu} =8 \beta^i \partial_i \tilde{a}
		= \chi_{ij}^g \beta^i \beta^j  \ ,
		\label{monot}
\end{equation}
which suggests viewing $\chi_{ij}^g$ as a metric in the space of couplings. 

Under the assumption that the metric $\chi_{ij}^g$ is positive definite  $\tilde{a}$ decreases  monotonically along the RG flow. The positivity of the metric $\chi_{ij}^g$ has been established at sufficiently small couplings to the highest known order in perturbation theory, and it  has been {\it conjectured} to hold non-perturbatively. Up to an arbitrariness in the definition of $ \chi^g $ (to be discussed later), this constitutes the {\it strong version} of the a-theorem {\it conjecture}, see for example \cite{Nakayama:2013is}.  In contrast, the {\it weak version} of the a-theorem has been {\it proven}~\cite{Komargodski:2011vj} and states that the quantity $\Delta {\tilde{a}} = ({\tilde{a}}_\text{UV} - {\tilde{a}}_{\text{IR}} ) = \Delta {a} = ({a}_\text{UV} - {a}_\text{IR}) > 0 $ for any RG flow between physical FPs, where $ \tilde{a}_{\mathrm{UV}(\mathrm{IR})} $ evaluated at the corresponding FPs.

Another consistency relation, derived with the same procedure as \eqref{eq:consistencycondition}, turns out to be particularly useful:
	\begin{equation}
	\label{eq:V_condition} 
	\chi^g_{ij} = - 2 \chi^a_{ij} + \bar{\chi}^{a}_{ijk} \beta^k - \beta^{\ell} \partial_\ell V_{ij} - \partial_i \beta^{\ell} V_{\ell j} - \partial_j \beta^{\ell} V_{i \ell}\ ,	
	\end{equation}
where 
	\begin{eqnarray}
	 \label{metricbara}
	V_{ij} &\equiv &\rho^{k} g^{k} \, \widebar{\mathcal{A}}_{ijk}^{(1)} \ , \qquad  \widebar{\mathcal{A}}_{ijk} \equiv \partial_k \mathcal{A}_{ij} - \tfrac{1}{2} \mathcal{B}_{ikj} - \tfrac{1}{2} \mathcal{B}_{jki}\ ,  \\  			
	\bar{\chi}^{a}_{ijk} &\equiv& \left(\epsilon - \hat{\beta}^{\ell} \partial_\ell \right) \widebar{\mathcal{A}}_{ijk} - \widebar{\mathcal{A}}_{\ell jk} \partial_i \hat{\beta}^{\ell} -  \widebar{\mathcal{A}}_{i\ell k} \partial_j \hat{\beta}^{\ell} - \widebar{\mathcal{A}}_{i j\ell} \partial_k \hat{\beta}^{\ell} \nonumber \ ,		
	\end{eqnarray}
and $\mathcal{A}_{ijk}^{(1)}$ is the residue of the $1/\epsilon$ pole. Thus, $\chi^g$ can be computed from CTs needed to renormalize contact divergences of marginal operators. 

In this Letter, we consider Yang-Mills theories with $N_f$ vector-like fermions in the $1/N_f$ expansion. We compute the metric $\chi_{ij}^g$ and the a-function $\tilde{a}$ to LO in $1/N_f$ but to all orders in the gauge coupling. 

\section{Preliminaries to determine the metrics}
Using the finiteness of $ n $-point functions in the renormalized theory we show how to determine the CTs $ \mathcal{A}_{ij} $ and $ \mathcal{B}_{ijk} $ in regular flat space and constant $ g^{i} $. We start with the 2-point function 
	\begin{equation}\label{eq:finite_2point}
	\begin{split}
	\frac{\delta}{\delta g^{i}(x)} \frac{\delta }{\delta g^{j}(y)} W &= i \brakets{ [\mathcal{O}_i(x) ][\mathcal{O}_j(y) ]} + \brakets{\frac{\delta [\mathcal{O}_j (y)]}{\delta g^{i}(x)}},
	\end{split}
	\end{equation}
which must be finite. Working with the flat metric $ \gamma_{\mu\nu} = \eta_{\mu\nu} $ and following~\cite{Jack:1990eb}, we define
	\begin{equation}
	[\mathcal{O}_i]^c \equiv \left. [\mathcal{O}_i] \right|_{\partial_\mu g=0} = \partial_i h^{a} \mathcal{O}_a^{0} \ .
	\end{equation}
The last equality embodies that the standard constant-coupling operators $ [\mathcal{O}_i]^c $ can always be expanded in terms of some functions $ h^{a}(g^{i}) $ and coupling independent operators $ \mathcal{O}^0_{a} $. 
Only the CTs in $ \widetilde{{\cal L}}_0 $ have dependence on $ \partial_\mu g^{i} $, so when the limit of spacetime-independent couplings is taken, we have
	\begin{eqnarray} \label{eq:derivative_of_operator}
	\dfrac{\delta [\mathcal{O}_j (y)]}{\delta g^{i}(x)} 
	&= K^k_{ij} [\mathcal{O}_k]^c \, \delta^d(x-y) + \mu^{-\epsilon} \mathcal{A}_{ij} \partial^4 \delta^d(x-y) \ , 
	\end{eqnarray}  
where $K^k_{ij} \equiv \partial_i \partial_j h^{a} \tfrac{\partial g^{k}}{\partial h^{a} }$. Also in this limit, the Fourier-transformed 2-point function is defined by 
	\begin{equation}
	\Gamma_{ij}(p) = i \int \dd^d x\, e^{-ip\cdot x} \brakets{ [\mathcal{O}_i(x)]^c [\mathcal{O}_j(0)]^c }\ .
	\end{equation}
The renormalized and finite 2-point function from \eqref{eq:finite_2point} takes the form
	\begin{equation}
	\Gamma^R_{ij}(p) = \Gamma_{ij}(p) + \mu^{-\epsilon} \mathcal{A}_{ij} (p^2)^2 +  K^k_{ij} \brakets{[\mathcal{O}_k]^c}\ .
	\label{2pointfinite}
	\end{equation}
This relation can be used to extract the $ \mathcal{A}_{ij} $ CT from the momentum-dependent part of $ \Gamma_{ij}(p) $ in flat space and constant couplings. 

Continuing on to the 3-point function, it is given by 
	\begin{equation} \label{eq:finite_3point}
	\begin{split}
	\frac{\delta}{\delta g^{i}(x)} &\frac{\delta}{\delta g^{j}(y)} \frac{\delta}{\delta g^{k}(z)} W  =  - \brakets{[\mathcal{O}_i(x) ] [\mathcal{O}_j(y) ] [\mathcal{O}_k(z) ]}  \\
	& + i\sum_{\mathrm{cyc}} \brakets{\frac{\delta [\mathcal{O}_j (y)]}{\delta g^{i}(x)} [ \mathcal{O}_k(z) ]} + i \brakets{\frac{\delta^2 [\mathcal{O}_k (z)] }{\delta g^{i}(x) \,\delta g^{j}(y) }} \ , 
	\end{split}
	\end{equation}
the sum being over cyclic permutations of $ i,j,k $ and $ x,y,z $.	
The Fourier-transformed 3-point function is defined by 
	\begin{equation}
	\Gamma_{ijk}(p, q) = - \int \dd^d x \dd^d y\, e^{-ip\cdot x} e^{-iq\cdot y} \brakets{ [\mathcal{O}_i(x)]^c [\mathcal{O}_j(y)]^c [\mathcal{O}_k(0)]^c}\ ,
	\end{equation}
to allow for computations in momentum space. 
The CTs of the renormalized 3-point function \eqref{eq:finite_3point} can be determined from the CTs \eqref{eq:RCT} and the relation \eqref{eq:derivative_of_operator}. Taking for simplicity $ \brakets{[\mathcal{O}_i]^c} =0 $, the finite 3-point function with constant couplings and flat metric is 
	\begin{align}
	\Gamma_{ijk}^R(p, q) &=  \Gamma_{ijk}(p, q) + K_{ij}^\ell \Gamma_{\ell k}(p +q) + K_{jk}^\ell \Gamma_{\ell i}(p) + K_{ki}^\ell \Gamma_{\ell j}(q) \nonumber \\
	&  + \mu^{-\epsilon} \left( \widebar{\mathcal{A}}_{ijk}\, p^2 q^2+ \widebar{\mathcal{A}}_{jki}\, q^2 (p+q)^2 + \widebar{\mathcal{A}}_{kij}\, (p+q)^2 p^2 \right) \nonumber \\
	& + \tfrac{1}{2}\mu^{-\epsilon} \left(\mathcal{B}_{ijk} (p+q)^4 + \mathcal{B}_{jki} \,p^4 + \mathcal{B}_{kij} \,q^4 \right)\ .		
	\label{3pointfinite}
	\end{align}
From here $ \bar{\mathcal{A}}_{ijk} $ can be extracted by, say, considering the term proportional to $ p^2 q^2 $, which does not receive contributions from any other CTs.
	
This sets up our computation of the $\mathcal{A}_{ij}$ and $\bar{\mathcal{A}}_{ijk}$ CTs which, through \eqref{eq:chi_a} and \eqref{eq:V_condition}, will allow us to obtain the metric $\chi^g_{ij}$.

\section{Large $N_f$  metric and  $ a $-function}
\begin{figure}[t]
	\includegraphics[width=.45\textwidth]{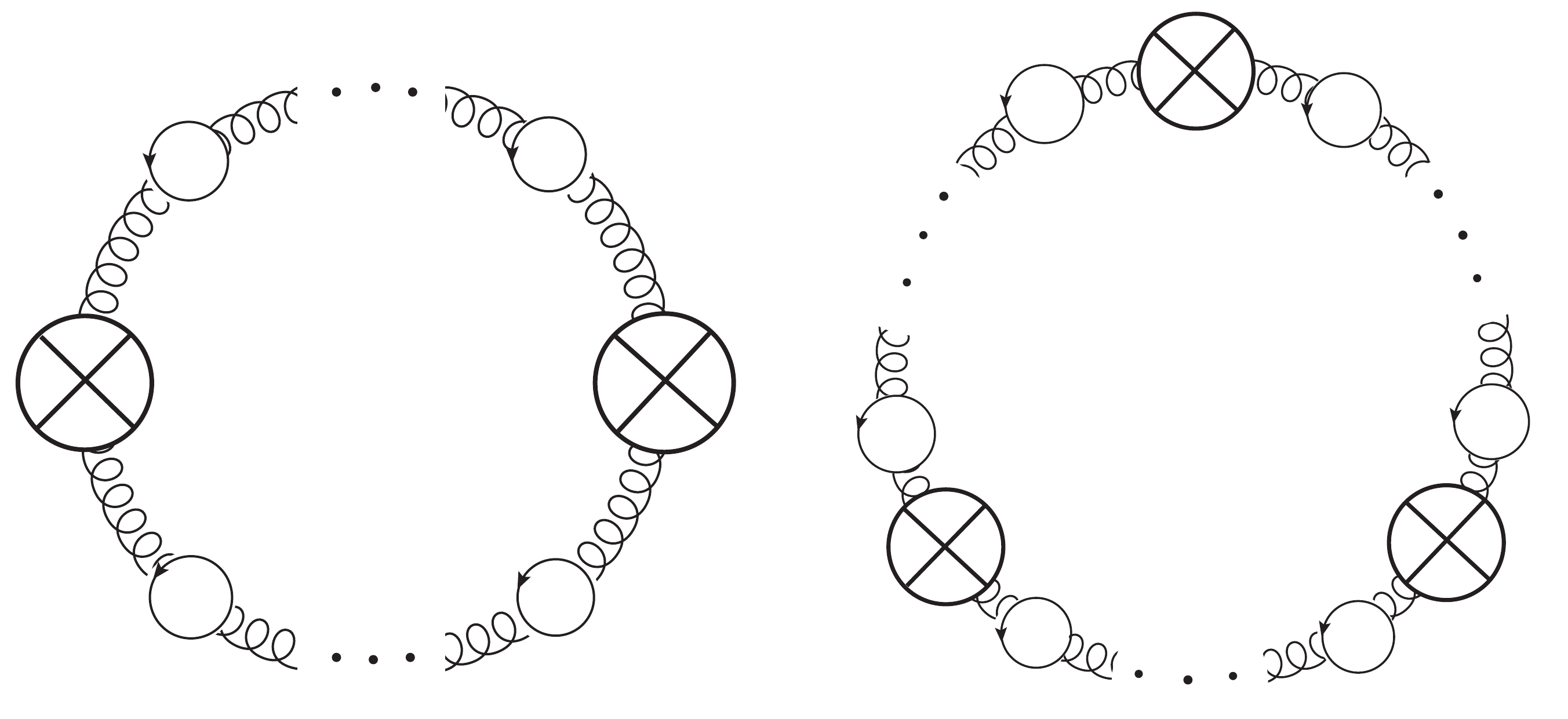} 
	\caption{LO 2- and 3-point functions. The crosses represent insertions of the $ [\mathcal{O}_g]^c $ composite operator. 
	}
	\label{metric}
\end{figure}
Consider a theory with large number of fermions, $ N_f $, charged under a simple (non-)Abelian gauge group;
	\begin{equation}
	\mathcal{L}_0 = -\dfrac{1}{4g_0^2} F_{0,\mu\nu}^{a} F_0^{a,\mu\nu} + \sum_{n=1}^{N_f} i \overline{\Psi}_{n,0} \gamma_\mu ( \partial^\mu - i A_0^\mu) \Psi_{n,0}\ .
	\end{equation}
Since the theory has a single (gauge) coupling we can suppress all coupling indices $i,j,k,...$ on the CTs. We now move to determine
the leading $ 1/N_f $ contribution to the 2- and 3-point correlation functions to all orders in perturbation theory, keeping the coupling $K\equiv g^2 N_f T_R/(4\pi^2)$ fixed  in order to prepare for the large $N_f$ limit. 

The leading 1PI correction to the gauge field 2-point function is given by the amputated fermion loop
	\begin{align} \label{eq:one_bubble_propagator}
	i \Pi_{\mu\nu} (p) &= i p^2 \Delta_{\mu\nu}(p) \mu^{-\epsilon} \, \Pi_0(p^2)\ , \nonumber \\
	\Pi_0(p^2) &= - \dfrac{N_f \,T_R}{2\pi^2} \dfrac{\Gamma^2(2-\tfrac{\epsilon}{2}) \Gamma(\tfrac{\epsilon}{2})}{\Gamma(4-\epsilon)} \left(-\dfrac{4\pi \mu^2}{p^2}\right)^{\epsilon/2} \ , 
	\end{align}
with the transverse projector $ \Delta_{\mu\nu}(p) = \eta_{\mu\nu} - p_\mu p_\nu/p^2$. One may then extract the LO contribution to the gauge field renormalization or equivalently to the coupling renormalization, setting 
	\begin{equation}
	K_0 = Z_A^{-1}\, K\ , \qquad Z_A = 1 - \dfrac{2 K}{3\epsilon} + \mathcal{O}(1/N_f)\ .
	\end{equation}
The beta function associated to the coupling is given by to LO in $ 1/N_f $
	\begin{equation} \label{eq:beta_K}
	\beta(K) = \tfrac{2}{3}K^2 + \mathcal{O}(1/N_f)
	\end{equation}	
	
Following this renormalization convention, the operator associated with $ K $ is found to be	
\begin{equation} \label{eq:Og}
	[\mathcal{O}_K]^c = \dfrac{N_f S_2(R)}{16 \pi^2 K^2 } \mu^{-\epsilon} F_0^2 + \mathcal{O}(1/N_f) \ . 
\end{equation}		
Only the $ F^2 $ term contributes at LO to the 2- and 3-point functions.  

At LO in the $1/N_f$ expansion, 2- and 3-point correlation functions are computed by dressing the gluon propagators with fermion bubble chains as shown in Fig.\ref{metric}. For the 2-point function, we have
	\begin{align}
	\Gamma_{KK}(p) =& \dfrac{ i \,d(G)}{2 K^2 Z_A^{2} } \int \dfrac{\dd^d k}{(2\pi)^d} 
	\dfrac{-i \Delta^{\mu\nu}(k) }{k^2 [1 - \Pi_0(k)]} V_{\nu\rho}(k,k + p) \nonumber \\
	& \quad \times \dfrac{-i \Delta^{\rho \sigma}(k+p) }{(k+p)^2 [1 - \Pi_0(k+p)]} V_{\sigma \mu}(k+p, k)\ ,
	\end{align} 
where $ V_{\mu\nu}(p,q) = p\cdot q\,  \eta _{\mu\nu} - q_\mu p_\nu $, is the momentum dependent Feynman rule stemming from a $ F^2 $-vertex and $d(G)\equiv N_c^2-1$ is the number of gauge bosons in the loop. Here we summed over every number of bubble insertions, which for the purpose of finding the divergent part is equivalent to using dressed gauge propagators. 
Extracting the $ (p^2)^2 $ dependent part of $ \Gamma_{KK}(p) $ and setting $ p\rightarrow 0 $, the integral may be evaluated using elementary methods. The $ \mathcal{A} $ CT is then determined from the finiteness of the renormalized 2-point function \eqref{2pointfinite} to be
	\begin{align}
	\label{eq:A_counter_term}
	\mathcal{A} &=  - \dfrac{3\, d(G)}{64 \pi^2 K^2 } \div \dfrac{H^{a}(\epsilon)}{Z_A^{2} K_0}\ , \qquad \text{where} \\
H^{a}(x) &= \dfrac{(1 - \tfrac{x}{3} )(240 - 240 x + 90 x^2 - 15 x^3 +x^4 ) \Gamma(4-x)}{60 (4-x) (6 - x) \Gamma(1 + \tfrac{x}{2}) \Gamma^3(2 - \tfrac{x}{2})} \ . \nonumber
	\end{align}
Here `$ \div $' is taken to mean the divergent part of the expression as $ \epsilon \rightarrow 0 $. 
	
Similarly, we evaluate the divergent part of the $ p^2 q^2 $ term in the 3-point function $ \Gamma_{KKK}(p,q) $ \eqref{3pointfinite} which allows us to determine the $\bar{\mathcal{A}}$ CT in \eqref{metricbara}. We find
	\begin{align}
		\label{eq:Abar_counter_term}
		\widebar{\mathcal{A}} &= \dfrac{d(G) }{64 \pi^2 K^3} \div \dfrac{\widebar{H}^{a}(\epsilon) }{Z_A^3 K_0} \ ,\qquad \text{where} \\
		\widebar{H}^{a}(x)&=  \dfrac{(80 - 60x +13 x^2 - x^3) x \, \Gamma(4-x) }{120 (4-x) \Gamma(1 + \tfrac{x}{2}) \Gamma^3(2 - \tfrac{x}{2})} \ . \nonumber
	\end{align}

The $ 1/\epsilon $ pole of $ \mathcal{A} $ and $ \bar{ \mathcal{A}} $ can be extracted using $ Z_A K_0 = K $. Since both $ H^{a} $ and $ \bar{H}^{a} $ are regular they are expanded as power series and then resummed in the coupling $ K $ at the simple pole in a manner similar to~\cite{PalanquesMestre:1983zy}. Inserting these results back into Eqs. \eqref{eq:chi_a}, \eqref{eq:V_condition}, and \eqref{metricbara} yields 
	\begin{align}
	\chi^{a}  &= -\tfrac{d(G)}{32 \pi^2 K^2} \partial_K \left[ K\,  H^{a}(\tfrac{2}{3} K) \right] 
	= -\tfrac{d(G)}{32 \pi^2 K^2} \left(1 - \tfrac{5}{3} K  + \tfrac{49}{108} K^2 + \ldots  \right) \ , \nonumber \\
	\chi^g &= \tfrac{d(G)}{16 \pi^2 K^2} \partial_K \left[ K\, H^{a}(\tfrac{2}{3} K) - \tfrac{1}{9} K^2 \,\widebar{H}^{a}(\tfrac{2}{3}K) \right] \nonumber \label{metricg}\\
	&=\tfrac{d(G)}{16 \pi^2 K^2} \left(1 - \tfrac{5}{3} K  + \tfrac{25}{108} K^2 + \ldots  \right)\ . 
	\end{align}	
Our result for $\chi^{a}$ agrees with  
\cite{Jamin:2012nn} to all-orders
and to $\mathcal{O}(K^2)$ with \cite{Zoller:2016iam}. Both metrics also agree with \cite{Jack:1990eb} to $\mathcal{O}(K)$.
Notice that the LO result only distinguishes the Abelian and non-Abelian theory through an overall normalization because gauge self-interactions are subleading in $1/N_f$. 
The theories also share the LO in $ 1/N_f $ beta function, Eq. \eqref{eq:beta_K}, with corresponding Landau pole. 
Using \eqref{eq:consistencycondition}, we can now derive the LO a-function (the contribution from the one-form $w_i$ vanishes)
	\begin{equation}
	\tilde{a} - \tilde{a}_{\text{free}} = \int \dd K \, \frac{K^2}{12} \chi^g 
	= \dfrac{d(G)}{192 \pi^2} \left[ K H^{a}(\tfrac{2}{3} K) - \tfrac{1}{9} K^2 \bar{H}^{a}(\tfrac{2}{3}K) \right] \ ,
	\end{equation}
where $ K $ is the large $ N_f $ coupling defined previously and $\tilde{a}_{\text{free}}$ is the free field theory result:
\begin{equation}
	\tilde{a}_{\text{free}} = \frac{1}{90 (8\pi)^2} \left( 11 N_f + 62 \, d(G) \right) \approx \frac{11 N_f}{90 (8\pi)^2} \ .
	\label{eq:afree}
\end{equation}

\begin{figure}
\hspace{-.03\textwidth}\includegraphics[width=.49\textwidth]{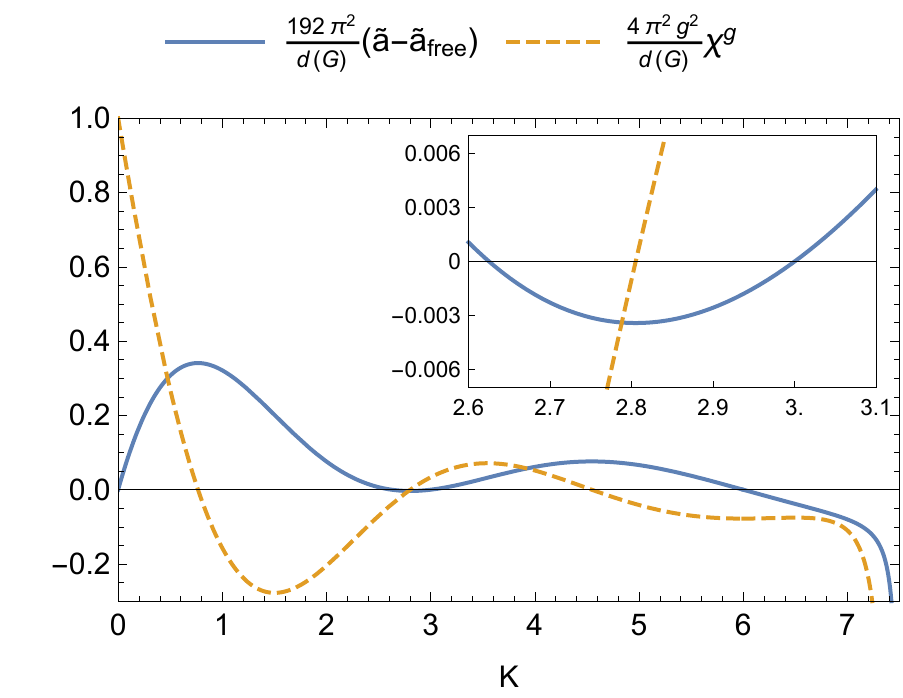} 
\vspace{-1em}
\caption{The LO in $ 1/N_f $ metric and $ a $-function.}
\label{Plotmetric}
\end{figure}

\begin{figure}
	\hspace{-.03\textwidth}\includegraphics[width=.49\textwidth]{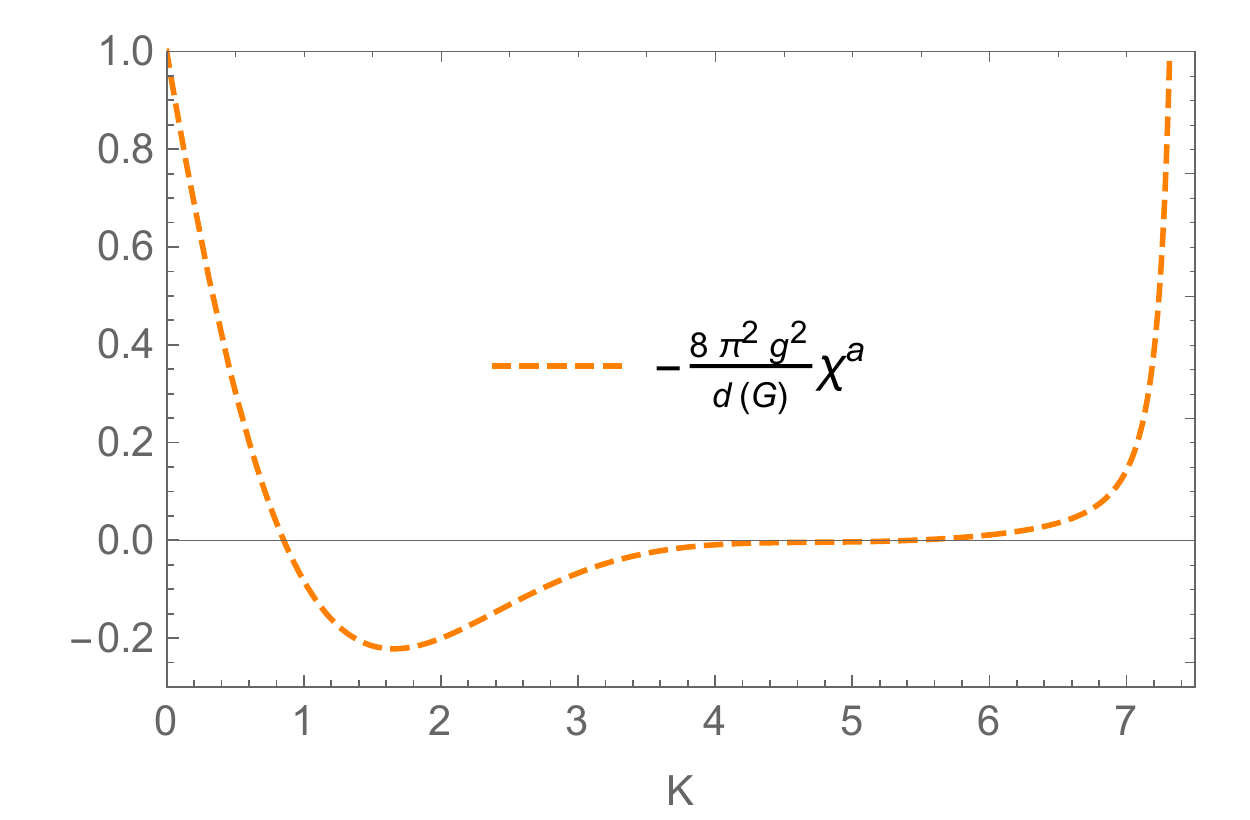} 
	\vspace{-1em}
	\caption{The LO in $ 1/N_f $ function $\chi^a$.}
	\label{Plotmetrica}
\end{figure}

Finally, in Fig.~\ref{Plotmetric} we plot the metric $\chi^g $ and
the a-function $\tilde{a}  - \tilde{a}_{\text{free}}$. We conclude that the metric is not positive definite for all values of $K$ and thus, the a-function is not monotonic, violating the strong version of the $a-$theorem. For completeness, the function $\chi^a$ is plotted in Fig.~\ref{Plotmetrica}.

One natural interpretation to restore the strong version of the $a$-theorem is that the flow towards the IR should start at $K$ no higher than $ \approx 0.8$ in the UV.  In fact, to this order in $ 1/N_f $, the underlying theory is UV incomplete and should be considered as an effective field theory that could be trusted  up to a maximum value of the energy corresponding to $K\approx 0.8$. In other words monotonicity of the $a$-function gives us a sense of how far in the UV the theory can be pushed as an effective field theory. As the underlying theory is UV incomplete, we cannot impose the weak version of the $a$-theorem. We notice however that if we were to impose it, we can extend the validity of the theory to $K\approx  2.6$ where the $a$-function becomes slightly negative. In the interval  $3 < K < 6$ we find that $a$ is positive albeit not monotonic. In between $K > 6$ and  $K=15/2$ we find that $a$ is negative with the LO metric and the $a$-function having a pole at $K=15/2$.

Different versions of the $a$-function can be obtained by redefining $ a $ with an arbitrary function $f(K)$ parameterizing the RG scheme change~\cite{Jack:2013sha}:
$\tilde{a}^\prime = \tilde{a} + f(K) \beta^2$ and simultaneously modifying the metric to $\chi^{g\prime} = \chi^g+8 [\beta \partial_K f(K)+2 f(K) \partial_K \beta]$ so that \eqref{monot} is invariant. To LO in $1/N_f$, we have:
\begin{equation}
\tilde{a}^\prime= \tilde{a}+\tfrac{4K^4f(K)}{9}\quad \chi^{g\prime}  = \chi^g +16 K \left[\tfrac{K\partial f(K)}{3\partial K} +\tfrac{4}{3} f(K)\right]  \ . 
\label{eq:redundancy}
\end{equation} 
Other proposals have been used in e.g. \cite{Komargodski:2011vj,Luty:2012ww,Prochazka:2018wkn}. It is not known if other versions of the $ a $-function are monotonic outside perturbation theory. Nevertheless, we have shown that the Jack and Osborn version~\cite{Jack:1990eb} is not monotonic.

\section{Subleading corrections and outlook}
At LO in $1/N_f$, no UV fixed point can emerge and the theory is therefore at best viewed as an effective field theory. However, the situation becomes intriguing upon considering the $1/N_f$ corrections. In particular, it has been argued in favor of the existence of an interacting UVFP for gauge-fermion theories due to the interplay between the leading and the subleading terms in $1/N_f$. Although the result is not as well established as the discovery of asymptotic safety in four dimensions in the Veneziano-Witten limit \cite{Litim:2014uca}, it has nevertheless led to a number of phenomenological \cite{Abel:2017rwl,Mann:2017wzh,Pelaggi:2017abg} and theoretical investigations \cite{Shrock:2013cca,Antipin:2018asc} culminating in the conformal window 2.0 \cite{Antipin:2017ebo}. According to the studies above, the UVFP for the fundamental representation occurs at 
\begin{eqnarray}
\label{Kstar}
K^*_{QCD}&=&4 N_f T_R\alpha^*=3-\exp \big[-p \frac{N_f}{N_c}+k\big] \ ,\nonumber\\ 
K^*_{QED}&=&4 N_f \alpha^*=\frac{15}{2}- 0.0117e^{-15\pi^2N_f/7} \ ,
\end{eqnarray}
where $p=16 T_R$ and $k=15.86+2.63/N_c^2$. The UVFP is expected to appear above some critical number of flavors $N_f^\mathrm{crit}$ above which the large $N_f$ expansion is reliable.  We notice that, for both Abelian and non-Abelian theories, the UVFP occurs for $K>$ 0.8 and thus the strong version of the a-theorem is necessarily violated. Intriguingly, the non-abelian fixed point occurs very close to the apparent loss of validity of the weak $a$-theorem. Of course, very near to the UVFP one must include the missing $1/N_f$ corrections to the $a$-function. For the Abelian case the alleged UVFP occurs at a pole of the $a$-function.  

If an UVFP exists for the non-Abelian case, only the weak version of the $a$-theorem survives, the reason being that the $1/N_f$ corrections cannot change the non-monotonic character of the $a$-function away from the UVFP. 

We elucidated the dynamics of large $N_f$ gauge-fermion theories by determining important properties such as the metric and $a$-function, for the first time, to all orders in perturbation theory. Our results can be tested via first principle lattice simulations, and can be further extended to multiple-couplings theories at large $N_f$ \cite{Kowalska:2017pkt,Antipin:2018zdg,Alanne:2018ene}. 

We thank Colin Poole and Vladimir Prochazka for discussion and helpful comments.
The work of O.A. is partially supported by the Croatian Science Foundation project number 4418 as well as the H2020 CSA Twinning project No.692194, RBI- T-WINNING while N.D., F.S., and A.E.T. are partially supported by the Danish National Research Foundation grant DNRF:90.


\begin{thebibliography}{99}

\bibitem{Cardy:1988cwa}
  J.~L.~Cardy,
  Phys.\ Lett.\ B {\bf 215} (1988) 749.
  doi:10.1016/0370-2693(88)90054-8
  
\bibitem{Zamolodchikov:1986gt}
  A.~B.~Zamolodchikov,
  JETP Lett.\  {\bf 43} (1986) 730
   [Pisma Zh.\ Eksp.\ Teor.\ Fiz.\  {\bf 43} (1986) 565].
  
\bibitem{Holdom:2010qs}
B.~Holdom,
Phys.\ Lett.\ B {\bf 694} (2011) 74
doi:10.1016/j.physletb.2010.09.037
[arXiv:1006.2119 [hep-ph]].

\bibitem{Jack:1990eb}
  I.~Jack and H.~Osborn,
  Nucl.\ Phys.\ B {\bf 343} (1990) 647.
  doi:10.1016/0550-3213(90)90584-Z

\bibitem{Osborn:1991gm}
H.~Osborn,
Nucl.\ Phys.\ B {\bf 363} (1991) 486.
doi:10.1016/0550-3213(91)80030-P

\bibitem{Baume:2014rla}
F.~Baume, B.~Keren-Zur, R.~Rattazzi and L.~Vitale,
JHEP {\bf 1408} (2014) 152
doi:10.1007/JHEP08(2014)152
[arXiv:1401.5983 [hep-th]].

\bibitem{Jack:2013sha}
I.~Jack and H.~Osborn,
Nucl.\ Phys.\ B {\bf 883} (2014) 425
doi:10.1016/j.nuclphysb.2014.03.018
[arXiv:1312.0428 [hep-th]].

\bibitem{Nakayama:2013is}
  Y.~Nakayama,
  Phys.\ Rept.\  {\bf 569} (2015) 1
  doi:10.1016/j.physrep.2014.12.003
  [arXiv:1302.0884 [hep-th]].
  
\bibitem{Komargodski:2011vj}
  Z.~Komargodski and A.~Schwimmer,
  JHEP {\bf 1112} (2011) 099
  doi:10.1007/JHEP12(2011)099
  [arXiv:1107.3987 [hep-th]].
  
\bibitem{PalanquesMestre:1983zy}
A.~Palanques-Mestre and P.~Pascual,
Commun.\ Math.\ Phys.\  {\bf 95} (1984) 277.
doi:10.1007/BF01212398

%
%
%
%
\bibitem{Jamin:2012nn}
  M.~Jamin,
  JHEP {\bf 1204} (2012) 099
  doi:10.1007/JHEP04(2012)099
  [arXiv:1202.1169 [hep-ph]].
  
\bibitem{Zoller:2016iam}
  M.~F.~Zoller,
  JHEP {\bf 1604} (2016) 165
  doi:10.1007/JHEP04(2016)165
  [arXiv:1601.08094 [hep-ph]].

\bibitem{Litim:2014uca}
  D.~F.~Litim and F.~Sannino,
  JHEP {\bf 1412} (2014) 178
  doi:10.1007/JHEP12(2014)178
  [arXiv:1406.2337 [hep-th]].
  
\bibitem{Abel:2017rwl} 
  S.~Abel and F.~Sannino,
  Phys.\ Rev.\ D {\bf 96}, no. 5, 055021 (2017)
  doi:10.1103/PhysRevD.96.055021
  [arXiv:1707.06638 [hep-ph]].

\bibitem{Luty:2012ww}
M.~A.~Luty, J.~Polchinski and R.~Rattazzi,
JHEP {\bf 1301} (2013) 152
doi:10.1007/JHEP01(2013)152
[arXiv:1204.5221 [hep-th]].

\bibitem{Mann:2017wzh} 
  R.~Mann, J.~Meffe, F.~Sannino, T.~Steele, Z.~W.~Wang and C.~Zhang,
  Phys.\ Rev.\ Lett.\  {\bf 119}, no. 26, 261802 (2017)
  doi:10.1103/PhysRevLett.119.261802
  [arXiv:1707.02942 [hep-ph]].
  
\bibitem{Pelaggi:2017abg} 
  G.~M.~Pelaggi, A.~D.~Plascencia, A.~Salvio, F.~Sannino, J.~Smirnov and A.~Strumia,
  Phys.\ Rev.\ D {\bf 97}, no. 9, 095013 (2018)
  doi:10.1103/PhysRevD.97.095013
  [arXiv:1708.00437 [hep-ph]].
  
\bibitem{Shrock:2013cca}
  R.~Shrock,
  Phys.\ Rev.\ D {\bf 89} (2014) no.4,  045019
  doi:10.1103/PhysRevD.89.045019
  [arXiv:1311.5268 [hep-th]].

\bibitem{Antipin:2018asc}
  O.~Antipin, A.~Maiezza and J.~C.~Vasquez,
  arXiv:1807.05060 [hep-th].

\bibitem{Antipin:2017ebo}
  O.~Antipin and F.~Sannino,
  Phys.\ Rev.\ D {\bf 97} (2018) no.11,  116007
  doi:10.1103/PhysRevD.97.116007
  [arXiv:1709.02354 [hep-ph]].

\bibitem{Antipin:2018zdg}
  O.~Antipin, N.~A.~Dondi, F.~Sannino, A.~E.~Thomsen and Z.~W.~Wang,
  Phys.\ Rev.\ D {\bf 98} (2018) no.1,  016003
  doi:10.1103/PhysRevD.98.016003
  [arXiv:1803.09770 [hep-ph]].

\bibitem{Kowalska:2017pkt}
  K.~Kowalska and E.~M.~Sessolo,
  JHEP {\bf 1804} (2018) 027
  doi:10.1007/JHEP04(2018)027
  [arXiv:1712.06859 [hep-ph]].
  
\bibitem{Alanne:2018ene}
  T.~Alanne and S.~Blasi,
  arXiv:1806.06954 [hep-ph].
  
\bibitem{Prochazka:2018wkn}
  V.~Prochazka and R.~Zwicky,
  arXiv:1807.06915 [hep-th].
  
\end{thebibliography}
\end{document}